\documentclass {article}
\usepackage{spconf,amsmath,graphicx,float}
\usepackage{url,subfigure,cite,array}
\usepackage{hyperref} 
\usepackage{color}
\usepackage{booktabs}
\usepackage{tabularx}
\usepackage{multirow}
\title{One-shot voice conversion for style transfer based on speaker adaptation}
\name{Zhichao Wang$^1$, Qicong Xie$^1$, Tao Li$^1$, Hongqiang Du$^1$, Lei Xie$^1$$^*$, Pengcheng Zhu$^2$, Mengxiao Bi$^2$}

\address{
  $^1$Audio, Speech and Language Processing Group (ASLP@NPU)\\School of Computer Science,
  Northwestern Polytechnical University, Xi’an, China\\
  $^2$Fuxi AI Lab, NetEase Inc., Hangzhou, China}
\begin{document}
\ninept
\maketitle
\begin{abstract}
One-shot style transfer is a challenging task, since training on one utterance makes model extremely easy to over-fit to training data and causes low speaker similarity and lack of expressiveness. In this paper, we build on the recognition-synthesis framework and propose a one-shot voice conversion approach for style transfer based on speaker adaptation. First, a speaker normalization module is adopted to remove speaker-related information in bottleneck features extracted by ASR. Second, we adopt weight regularization in the adaptation process to prevent over-fitting caused by using only one utterance from target speaker as training data. Finally, to comprehensively decouple the speech factors, i.e., content, speaker, style, and transfer source style to the target, a prosody module is used to extract prosody representation. Experiments show that our approach is superior to the state-of-the-art one-shot VC systems in terms of style and speaker similarity; additionally, our approach also maintains good speech quality.
\end{abstract}

\renewcommand{\thefootnote}{\fnsymbol{footnote}}
\footnotetext{*Corresponding author.}

\vspace{-6pt}
\section{Introduction}

\begin{figure*}[h]
  \centering
  \includegraphics[width=0.9\linewidth]{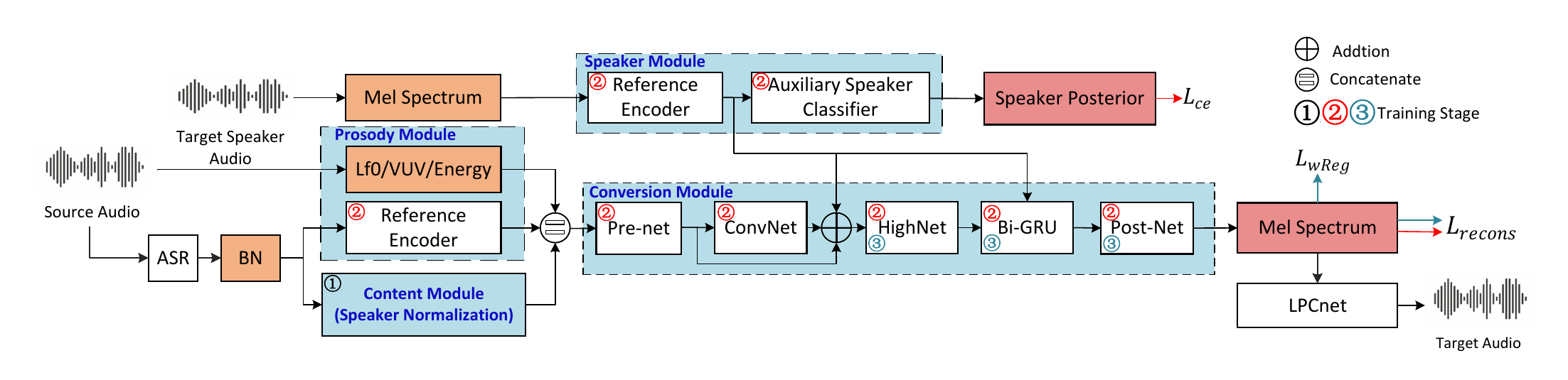}
  \vspace{-10pt}
  \caption{Overview of the components of the proposed system.}
  \label{fig:model_framework}
\end{figure*}

Voice conversion (VC) aims to modify speech from a source speaker to sound like that of a target speaker without changing the linguistic content. Voice conversion based on neural networks, e.g. GAN~\cite{GANHsu2017VoiceCF}, VAE~\cite{VAEHsu2016Voicevae}, recognition-synthesis\cite{PPGSun2016PhoneticPF} frameworks, have significantly improved speech quality and speaker similarity. Despite recent progress, most current works focus on the transformation of timbre while ignoring the transformation of style contained in source speech. Recently, explicit prosodic features~\cite{explicitMing2016ExemplarbasedSR,explicitRaitio2020prosodyControl} and implicit prosody extractor~\cite{impiicitLiu2020TransferringSS,EnrichingSourceStyleTransfer} are used to model prosody and perform source style transfer, while requiring a large number of recordings. Data collection is expensive and time-consuming. Therefore, it remains a challenge to build a high-quality voice conversion framework for style transfer with limited data.

Recently, one-shot voice conversion approaches are proposed, which refers to the conversion of the speaker identity given only one utterance of the target speaker, which is more practical for application as compared with previous methods in which sizeable data for the target speaker is needed. Some studies~\cite{autovcqian2019autovc,AgainVC,OSSEDu2021ImprovingRO} propose to extract content and speaker representation from speech during training stage, and concatenate content representation from source speech and speaker representation from target speech at run-time to produce converted speech. However, the content representation may still contain speaker-related information, which results in unstable performance in terms of speaker similarity. Other works~\cite{GSE,VQMIVC} decompose speech into three factors: content, speaker, and pitch with the premise that pitch is highly related to speaker characteristics. These methods further improve the disentanglement ability, while there is still a gap between the converted speech and target speech in terms of speaker similarity. To improve speaker similarity, many works~\cite{Adaspeech,TTSADaptationArik2018NeuralVC,GCtts} in text to speech (TTS), a related task, focus on few-shot speaker adaptation. Unlike few-shot ($\geq$10 utterances) TTS, in one-shot voice conversion, training on one utterance (3-4s) can easily lead to over-fitting and reduce the speaker similarity and speech quality. Besides delivering correct content and speaker information, transferring style or prosody from source speaker to the target is also desired, but more challenging~\cite{explicitMing2016ExemplarbasedSR,explicitRaitio2020prosodyControl,impiicitLiu2020TransferringSS,EnrichingSourceStyleTransfer}, as prosodic aspects are entangled with content and speaker as well and the over-fit problem is rather complicated. 

\vspace{-1pt}
In our paper, we address the \textit{one-shot style transfer} problem for voice conversion, aiming to transferring the source content and style to the target speaker while maintaining the good target speaker identity; more challengingly, we achieve the above goal with only one utterance with several seconds from the target speaker. Specifically, to solve the over-fit problem caused by training with only one utterance, we propose a novel one-shot voice conversion framework for style transfer based on the recognition-synthesis framework~\cite{PPGSun2016PhoneticPF}, integrating speaker normalization~\cite{speakernormalzationGlarner2021}, weight regularization~\cite{weightRegLi2020}, and prosody modeling. Specifically, speaker normalization is used to remove speaker-related information in the bottleneck feature (BN) extracted from ASR, which is helpful to improve speaker similarity. Weight regularization is applied to the adaptation with one utterance to prevent performance degradation. A prosody module~\cite{EnrichingSourceStyleTransfer} is adopted to explicitly extract prosody information for transfer. We believe such a comprehensive characterization of speech factors can achieve good style transfer and make the model focus on learning timbre in the adaptation process. The experimental results on VCTK~\cite{VCTK} and CMU-ARCTIC~\cite{CMU-Arctic} show that over-fitting is greatly alleviated after using these methods, and our framework is better than several state-of-the-art (SOTA) methods including AGAINVC~\cite{AgainVC}, GSE~\cite{GSE} and VQMIVC~\cite{VQMIVC} in terms of style and speaker similarity.

% \vspace{-1pt}
The rest of the paper is organized as follows. Section 2 introduces our proposed methods. Section 3 presents the experiments and compares our proposed system with SOTA systems with subjective and objective evaluation. Section 5 concludes this paper.

\vspace{-4pt}
\section{Proposed approach}
\vspace{-3pt}
\subsection{System Overview}
\vspace{-3pt}

Our proposed one-shot voice conversion framework is based on a recognition-synthesis architecture, which is shown in Fig.~\ref{fig:model_framework}. This framework consists of four main modules (light blue boxes in Fig.~\ref{fig:model_framework}): content module, speaker module, prosody module and conversion module. The content module takes bottleneck features (BN) extracted from ASR as input, and the output is speaker-independent content representation. Specifically, we adopt speaker normalization technique to remove speaker-related information contained in BN. The speaker module takes mel spectrum from target speaker to extract the speaker representation. To ensure the speaker module is able to extract desired speaker representation without speaker confusion, we add an auxiliary speaker classifier after the reference encoder. The prosody module learns to extract speaker-independent prosody representations, which integrates explicit and implicit hybrid modeling methods~\cite{EnrichingSourceStyleTransfer}. The explicit modeling includes raw logarithmic domain fundamental frequency (lf0), normalized short-term average amplitude (energy), and the voice/unvoice flag (VUV). For implicit modeling, we adopt reference encoder to extract prosody representation. The conversion module takes prosody, content, and speaker representations as input, and the output is mel spectrum. Finally, modified LPCnet~\cite{LPCnet} is adopted to reconstruct waveform from mel spectrum.

\subsection{Speaker Normalization}
% 参考acoustic unit discorty 论文，用法一致，引用之前做对抗去说话人的思路，从输入上减轻模型的训练差异行
% 如图所示，我们使用说话人归一化模块将输入的任意的源音频的瓶颈特征特征转换成同一个特定说话人的语句作为转换模型的内容表征。一方面，受到对抗方法的启发\cite{ppgdvectoradv,AccentVCWang2020}，使用对抗训练的方式去除输入特征的说话人差异性防止影响最终的说话人相似度，另一方面，提出的瓶颈特征无法完全去除说话人相关的信息，比如音色和说话人相关的风格，需要更多模型的训练参数去学习拟合这种差异性。根据上述的两种现象，我们需要一个说话人归一化模块去除输入特征的说话人差异性，将模型的任务从any-to-one 转变成one-to-one。考虑到对抗性训练的训练成本和不稳定性，类似于Glarner et al.~\cite{speakernormalzationGlarner2021},我们使用完整的语音转换系统作为我们的说话人归一化模块，将任意说话人的不同的话语转换到同一个说话人上来达到去除content信息的说话人差异性

% As shown in Fig.~\ref{fig:model_framework}, \textcolor{red}{we adopt speaker normalization in content module to convert the source audio's BN into mel spectrum of the specific speaker as the content representation.}
Although an ASR system is trained using multi-speaker data, targeting to speaker-independent acoustic representation, studies~\cite{ppgdvectoradv,AccentVCWang2020,EnrichingSourceStyleTransfer} show that the bottleneck feature (BN) still inevitably contains speaker-related information, such as timbre and style. Thus for voice conversion, the speaker similarity of the converted speech may be degraded. As shown in Fig.~\ref{fig:model_framework}, we specifically introduce a speaker normalization method to remove the speaker-related information. Previously, adversarial training ~\cite{ppgdvectoradv,AccentVCWang2020} was usually used to achieve this goal. But taking into account the training cost and instability of adversarial training, we utilize a speaker normalization method~\cite{speakernormalzationGlarner2021} instead in our framework. Specifically, we utilize an any-to-one VC network here to serve as a speaker `normalization' trick to `normalize' the source audio's BN to mel spectrum of a specific speaker to achieve the purpose of normalizing content information. By doing so, we actually relieve the burden of the rest of our framework.

\subsection{Weight Regularization}
\label{sec:sn}
% 参考adapt，从过拟合的角度出发
% 如图所示，为了进一步缓解因为只在一句话上训练所带来的过拟合的问题同时提高训练过程的稳定性，在训练阶段3我们引入了权重正则的方法~\cite{weightRegLi2020}.使用权重正则loss LwReg prevent the
% parameters of the prediction model C to drift far away from those of the pre-trained model learnt in the source dataset。权重正则从计算方法看是l2正则的一个变种，可以有以下定义其计算方式（It can be defined as follows:）
% \begin{equation}
%     L_{wReg}=||\theta-\theta_{s}||^2
% \end{equation}
% 其中θ表示随着时间更新的训练参数，θs表示训练开始时模型的参数，同时整个过程中保持不动，在我们实验中，theta，thetas表示就是stage3的训练前后的模型参数。该正则化方法可以帮助模型训练过程参数不会变化过快导致的不稳定，同时帮助模型在训练之后仍然具有较高的转换质量。

As shown in Fig.~\ref{fig:model_framework}, in order to alleviate the over-fitting problem caused by training with only one sentence and improve the stability of the training process, we introduce weight regularization~\cite{weightRegLi2020}. The weight regularization is a variant of l2 regularization, defined as
\begin{equation}
    L_{wReg}=||\theta-\theta_{f}||^2,
\end{equation}
where $\theta$ represents the parameters of HighNet, Bi-GRU and postnet updated over time in adaptation process, and $\theta_{f}$ represents these three layers' parameters before adaptation which is used as fixed value during the adaptation process. The core idea in $L_{wReg}$ is to prevent the parameters of adapted model to drift far away from those of the base model learnt in the large dataset. The regularization prevents model from changing too significantly, which is helpful in improving the stability of the training process.

\subsection{Source Style Transfer}
% 引用之前的工作，从模型学习模式以及提高模型质量的角度说明
% 大多数之前的vc的one-shot工作都是基于在解耦content和speaker在重新组合这一模式上，其中 prosody这一关键信息被忽略 The prosody information at least includes emotion, pitch, duration, and loudness，这使韵律信息会泄露到content和speaker中降低解耦的稳定性，同时加大了模型的学习负担 which 模型不仅仅需要学习音色还需要学习韵律的转换在没有韵律模块的情况下。受到之前工作的启发【gse,enriching】，我们考虑将源风格转换这一任务引入框架，在content，prosody,speaker这一模式下完成解耦及组合。如之前的论文一致，我们设计了一个包含显隐式混合建模方法的韵律模块，不同点在于，显示建模中我们使用raw lf0，在推断是通过线性转换匹配目标说话人，希望能够通过显示的先验信息减轻模型训练负担。另外只使用了bn作为韵律模块隐式建模部分的输入。通过这一框架的引入，一方面能够通过增加更多的模型先验信息，使模型更加专注于对于目标说话人音色的学习，另外一方面，能够通过源风格转换的方法提高one-shot后语音的自然度和表现力。实验也能说明在引入框架后one-shot 自适应情况下模型的表现

Many one-shot voice conversion methods~\cite{autovcqian2019autovc,INchou2019oneshot,AgainVC,GSE,VQMIVC} focus on disentangling content and speaker, while ignoring the prosody modeling, which results in that the prosody information may leak to the content and speaker representation. This reduces the stability of disentanglement and causes poor speaker similarity, although the prosody of source is much transferred to the target. Inspired by our previous work~\cite{EnrichingSourceStyleTransfer}, we add a prosody module to the one-shot voice conversion framework. Recall that we explicitly decompose speech into three parts: content, speaker as well as style, and this prosody module is specifically in charge of the style representation for the final conversion module. In detail, the prosody module includes explicit and implicit modeling schemes. For explicit modeling, we extract energy, VUV, and raw lf0 from source audio. For implicit modeling, the BN is used as the input of a reference encoder, and the output is prosody representation. By adding the prosody module, the voice conversion framework is able to perform style transfer.

% \vspace{-10pt}
\begin{table*}[h]
\caption{MOS results with 95\% confidence interval.}
\vspace{10pt}
\label{tab:mos}
\centering
\begin{tabular}{c|l|l|ccc}
\toprule
\multicolumn{3}{c|}{Model}                                         & Speech Quality  & Style Similarity   & Speaker Similarity \\ \midrule
Comparison & \multicolumn{2}{l|}{AGAINVC~\cite{AgainVC}}        & 2.81$\pm$0.12          & 3.03$\pm$0.34          & 2.96$\pm$0.34          \\ \cline{2-6} 
                             & \multicolumn{2}{l|}{VQMIVC~\cite{VQMIVC}}         & 2.94$\pm$0.27          & 3.19$\pm$0.30          & 3.02$\pm$0.15          \\ \cline{2-6} 
                             & \multicolumn{2}{l|}{GSE~\cite{GSE}}            & \textbf{3.46$\pm$0.16} & 3.19$\pm$0.12          & 3.03$\pm$0.18          \\ \cline{2-6} 
                             & \multicolumn{2}{l|}{GSE-finetune} & 3.09$\pm$0.14           & 2.88$\pm$0.11          & 3.31$\pm$0.12          \\ \hline
Ablation    & BL  & Baseline                & 2.93$\pm$0.13           & 2.84$\pm$0.16          & 3.25$\pm$0.13          \\ \cline{2-6} 
                             & P1        & \ +Speaker Normalization  & 3.16$\pm$0.12           & 3.15$\pm$0.12          & 3.37$\pm$0.10          \\ \cline{2-6} 
                             & P2        & \ \ +Weight Regularization    & 3.27$\pm$0.12           & 3.21$\pm$0.13          & 3.43$\pm$0.07          \\ \hline
\multicolumn{2}{c|}{P3 (Proposed)}        & \ \ \ +Prosody Module         & 3.43$\pm$0.12           & \textbf{3.76$\pm$0.17} & \textbf{3.56$\pm$0.15} \\ \bottomrule
\end{tabular}
\end{table*}

\subsection{Training and Conversion Procedure}
% 说明各个步走中。1，2，3步骤中训练的模型，部分，作用以及为什么这么做

% As showns in Fig.~\ref{fig:model_framework},our approach is composed of three-stage training phase and a conversion phase. 我们遵循识别-合成框架，首先需要有预训练完成的ASR，这一部分未算在训练步骤中。正如图a所示，我们使用不同颜色的数字标注在我们框架中在训练的不同阶段需要参与更新的模型。我们将在下面详细说明：
% 1.训练阶段1: 说话人归一化模块使用大量说话人数据预训练基础模型增加模块稳定性，并在特定说话人上进行微调，以提高说话人归一化的效果。在我们实验中，说话人归一化模块使用mel reconstraction loss进行更新，或者也可以采用其他经典的语音转换框架实现。
% 2.训练阶段2：在该阶段中，prosody module, speaker module以及cbhg module参与模型训练。整个阶段在包含大量说话人的数据上进行训练。
% 对于Prosody module，我们提取瓶颈特征，原始对数域基频，归一化短时平均幅度谱以及voice/unvoice flag(vuv)作为模型的条件输入。
% 特别说明的是对于说话人模块，随机选取的同一个说话人的mel谱作为模块的输入，保证提取的说话人表征仅仅关注于说话人信息而与其他的信息无关比如文本信息，同时增加说话人分类器去增强说话人表征能力，说话人分类器训练使用交叉熵loss Lce进行更新.整个阶段的loss可以描述为：
% Lstage2=Lrecons+Lce
% 3.训练阶段3:这个阶段中，我们只有目标说话人的一句话来更新模型，类似于【boffin tts,adapaspeech,GCtts】，为了防止一句话训练模型的情况下，越大的参数量越容易导致过拟合，正如图中(a)标记3所示，我们冻结了prosody module,cbhg module的部分网络，同时为了防止训练好的说话人表征空间被扰乱，speaker encoder也同样被冻结。在冻结了这些部分之后，进一步为了防止过拟合导致的转换质量下降的现象产生，我们结合权重正则{权重论文}的方法保证在训练中网络权重不会过分偏离原来权重参数。整个阶段的loss总结如下
% Lstage3=Lrecons+LwReg

% 转换阶段如图（b）所示，从源音频提取bn and prodic features被提取作为转换模型的输入。与训练阶段不同的是，为了减小说话人之间的特征差异，lf0根据目标说话人的音高进行调节，运用线性转换：

% 公式
% 其中。。。。。，解释。同时，目标说话人的给定的一句话被用作参考音频提供说话人身份信息。最终模型根据文本，韵律，说话人表征将任意源音频转换成目标说话人的音频

As shown in Fig.~\ref{fig:model_framework}, our approach is composed of three training phases and a conversion phase. Different phases are marked with numbers in the figure.

\textbf{Training phase 1.} As we mentioned in Section~\ref{sec:sn}, we utilize an any-to-one VC method to implement our content module. The content module is trained on a large amount of speech data and finetunes on a specific speaker to ensure the performance of speaker normalization. In our experiment, the content module is optimized with mel reconstruction loss.

\textbf{Training phase 2.} The whole model is trained with a large amount of data. Note that the speaker module is trained with a randomly selected mel spectrum of the same speaker to make sure that the extracted speaker embedding is only related to the speaker information. The prosody and conversion modules are optimized with reconstruction loss $L_{recons}$. The speaker module is optimized with cross-entropy loss $L_{ce}$. The loss function of this stage can be described as follows:
    \begin{equation}
    Loss_{stage2}=L_{recons}+ L_{ce}
    \end{equation}

\textbf{Training phase 3.} This stage is for adaptation. In this stage, we only use one utterance from target speaker to fine-tune the model. Previous works~\cite{Adaspeech,GCtts} confirm that a large number of parameters may make model easy to over-fit. Hence, only part of the conversion module is involved in training, including HighNet, Bi-GRU and Post-Net, as shown in Fig.~\ref{fig:model_framework}. To further prevent performance degradation caused by over-fitting, we adopt the weight regularization~\cite{weightRegLi2020} in this phase. The loss used in this phase is described as follows:
    \begin{equation}
    Loss_{stage3}=L_{recons}+\gamma L_{wReg}
    \end{equation}

\textbf{Conversion phase.} The content and prosody representation are extracted from source speech. Note that we use linear transformation to obtain converted f0. The speaker representation is extracted from target speech. The conversion module takes content, prosody, and speaker representation to reconstruct converted speech. Through the proposed disentanglement and training procedure, the converted speech is expected to have the same prosody as source speech, while maintaining the content and target speaker's identity.

\section{EXPERIMENTS AND RESULTS}
\subsection{Dataset and Experimental Setup}

In our experiments, 102 speakers from VCTK~\cite{VCTK} are used to train the conversion model. For one-shot testing, \textit{p340}, \textit{p363} from VCTK, as well as \textit{slt}, \textit{bdl} from CMU-ARCTIC~\cite{CMU-Arctic} are used as the target speakers. The duration of target speech ranges from 3 to 4 seconds. For the source audio, we use CMU-ARCTIC (rms, clb) and ESD dataset~\cite{ESD}. All speech utterances are downsampled to 16kHz. We use 80-dim mel spectrum computed with 50ms frame length and 12.5ms frame shift. The ASR system is a TDNN-F model trained with LibriSpeech~\cite{panayotov2015librispeech} corpus containing 1k hours speech.
% The ASR system is a TDNN-F model trained with 1k hours standard English corpus. 
We use the 256-dim bottleneck features as the linguistic representation, which is extracted from the last fully-connected layer before softmax in the TDNN-F model. Modified LPCnet~\cite{LPCnet} is adopted to reconstruct waveform from mel spectrum. The vocoder is trained with speech data of 102 speakers from VCTK.

% We use a 'universal' melLPCnet vocoder, which is similar to LPCnet~\cite{LPCnet} but uses mel spectrum as input to generate waveform. The vocoder is trained using 102 speakers speech data from VCTK.

To validate our proposed method, we implement comparison and ablation systems. For comparison systems, we selected three SOTA one-shot VC methods, including AGAINVC~\cite{AgainVC}, GSE~\cite{GSE} and VQMIVC~\cite{VQMIVC}.  For a fair comparison, we finetune GSE  with target speaker's utterance, which is referred as GSE-finetune. For ablation analysis, we implement four systems BL, P1, P2, and P3 (proposed). Among them, BL is composed of speaker and conversion module to convert BN to mel spectrum. P1 uses speaker normalization based on BL, and P2 adopts weight regularization based on P1. P3 is our final proposed system combining all the contributions.

% \begin{itemize}
% \item\textbf{Comparison systems:} we selected three SOTA one-shot VC methods, including GSE~\cite{GSE}, VQMIVC~\cite{VQMIVC} and AGAINVC~\cite{AgainVC}. All three models are trained on VCTK, of which we use the official released model for the AGAINVC and VQMIVC.
% \item\textbf{GSE-Finetune:} since the previous one-shot methods are based on unseen speaker's embedding, for a fair comparison, we also selected GSE for finetuning.
% \item\textbf{BL:} baseline system consisting of CBHG module with prenet and postnet reconstructs mel spectrum directly from BN.
% \item\textbf{P1:} adopt speaker normalization module based on BL.
% \item\textbf{P2:} apply weight regularization to conversion model based on P1. 
% \item\textbf{P3:} the final system we proposed both as described in Section 2. P3 uses prosody module based on P2.
% \end{itemize}

For our proposed method, the content adopts the same model configuration as Tian \textit{et al.}~\cite{VCC2020_Tian2020}. The content module follows a typical encoder-decoder architecture using CBHG as the encoder and an auto-regressive module consisting of prenet, decoder RNN and postnet as the decoder. For the conversion module, the configuration of CBHG is the same as that of the content module. Prenet consists of 2 fully connected layers with 80 and 256 hidden units respectively. Postnet contains 4 1D-convolution layers with 3*3 kernel size and 256 filters and a fully connected layers with 80 hidden units. The architecture and hyper-parameters of the reference encoder follow the original configuration in \cite{Rerferceencoder}. This structure mainly includes 6 layers of convolution and GRU. Each convolution layer is composed of 3×3 filters with 2×2 stride, SAME padding, and ReLU activation. The number of filters in each layer is [32, 32, 64, 64, 128, 128]. Batch normalization is applied to every layer. The output of convolution layers is fed into GRU with 128 units. The speaker classifier consists of 3 fully connected layers. In training stage 2, the conversion model is trained for 120 epochs using batch size of 16. We use Adam optimizer to optimize our model with learning rate decay, which starts from 0.001 and decays every 20 epochs with decay rate 0.7. In training stage 3, we train the conversion model for 2000 steps using one utterance, and $\gamma$ is set to 1. The learning rate starts from 0.001 and decays every 200 steps in decay rate 0.5.

\begin{figure*}[h]
	 \vspace{-10pt}
	\centering
	\includegraphics[width=0.9\linewidth]{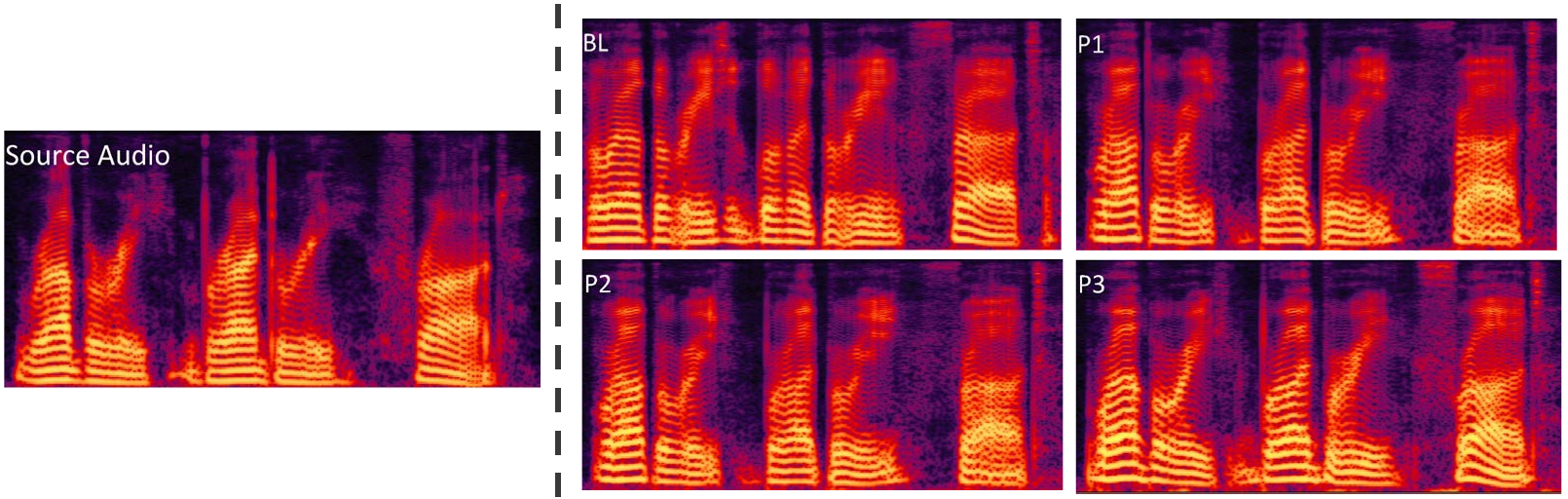}
\caption{Spectrograms for a source audio utterance (left) and its corresponding converted utterances by different systems (right). The formants from the proposed system (esp. P3) are more similar to the source, indicating good style transfer. By contrast, the formants from BL are too flat.}
\label{fig:mel}
\vspace{-6pt}
\end{figure*}

\subsection{Subjective evaluation}
We conduct the following listening tests: three mean opinion score (MOS) tests to assess speech quality, style similarity, and speaker similarity respectively. Style similarity measures styles between converted audio and source audio. We select three sentences from rms, clb, and ESD respectively as source audio. Nine sentences are converted to four target speakers (p340, p363, slt, bdl), a total of 36 sentences are used for listening test. We highly recommend readers to listen to our samples\footnote{Samples can be found in \href{https://kerwinchao.github.io/Oneshotvc.github.io/}{\url{https://kerwinchao.github.io/Oneshotvc.github.io/}}}.

% 1, which contains not only the results of this paper but also the results on the Chinese dataset.

\textbf{Comparison analysis.} We compare the proposed method with SOTA one-shot VC methods. The results of MOS tests are shown in Table~\ref{tab:mos} comparison part. It is observed that GSE achieves the best result in terms of speech quality, and our proposed system P3 gets significantly higher MOS scores in terms of style and speaker similarity than the comparison systems. Previous one-shot methods lack prosody modeling ability, and target speaker‘s timber is unknown to model, which leads to low speaker similarity and unstable performance. Comparing GSE with GSE-finetune, we can see that making GSE learn from the target speaker utterance can significantly improve the speaker similarity.

\textbf{Ablation analysis.} As shown in Table~\ref{tab:mos} ablation part, we evaluated the ablation systems. The BL system obtains poor results in three MOS tests, which indicates that BL system is easy to over-fit. By using our proposed method, the phenomenon of over-fitting is alleviated, and all MOS scores are improved. Adding the prosody module makes the speaker module focus on extracting timber from only one utterance, therefore that style and speaker similarity are able to be improved.

\begin{figure}[h]
	%  \vspace{-10pt}
	\centering
	\includegraphics[width=0.9\linewidth]{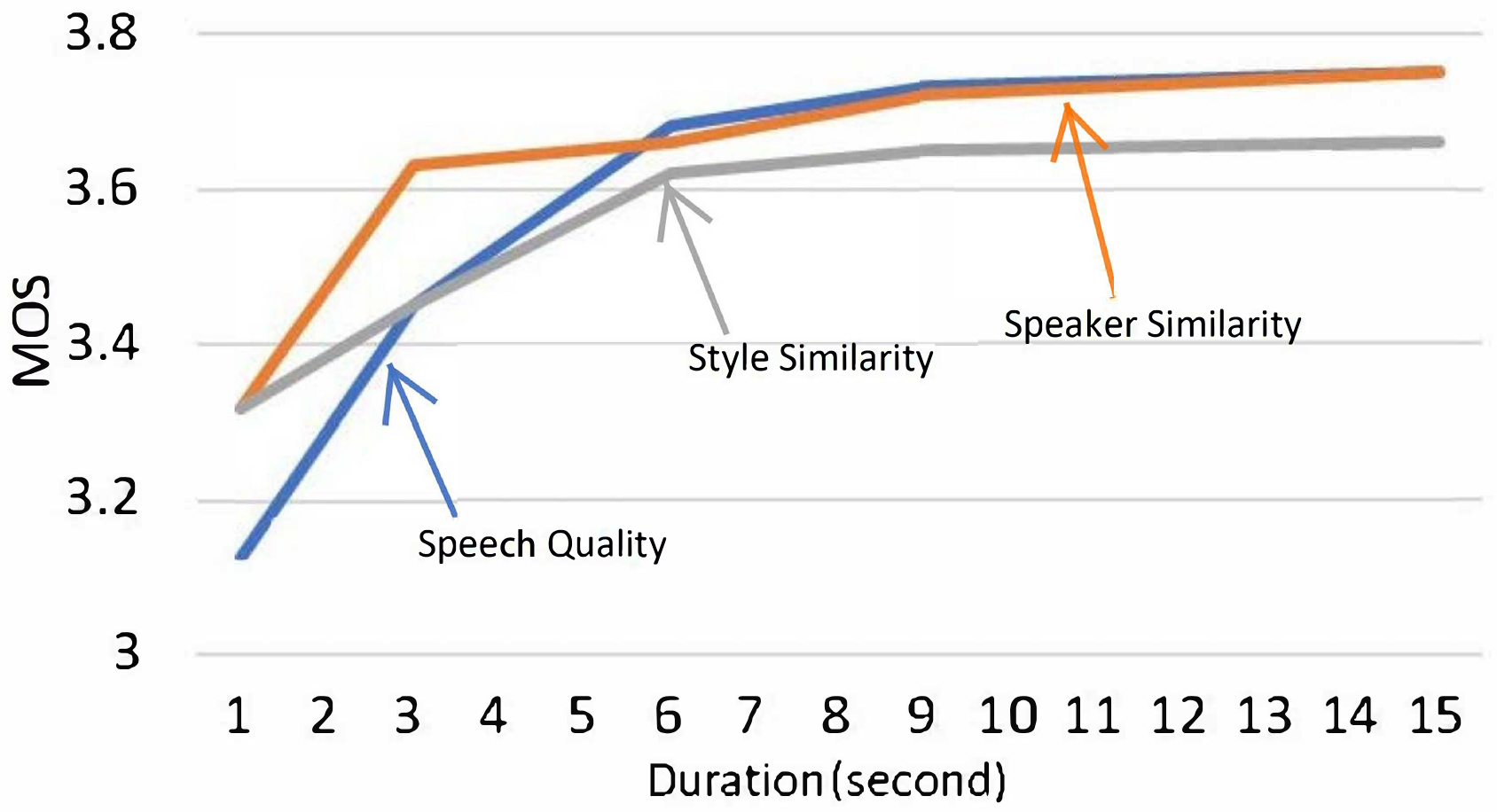}
	\caption{The MOS performance of proposed system vs. training utterance with different duration}
	\label{fig:duration}
\end{figure}

\textbf{Varying duration.} We further evaluate the performance of the proposed system under different duration of utterances (1, 3, 6, 9, 15 seconds) from CMU-ARCTIC speakers. Note that here only CMU-ARCTIC speakers are used for subjective test as listeners need to assess a large set of audio samples. As shown in Fig.~\ref{fig:duration}, the result is affected by the duration of the target speech in general. For the extreme cases, e.g. 1-3s, the model still shows quality degradation caused by over-fitting even if our proposed method is used. But from 3 to 6s, benefiting from the proposed approach, the three MOS scores have clear increased while speech quality and style similarity undergo a quick boost. After 6 seconds, all three curves do not improve significantly and begin to stabilize from 9 seconds.

\subsection{Objective evaluation}
% \vspace{-5pt}
\textbf{Over-fitting on spectrograms}. We visualize the spectrogram of the converted speech for further investigation. The spectrogram of a testing sample is shown in Figure~\ref{fig:mel}. We can see that BL has a very flat formant on the spectrogram, and noise appears in the silence part. Comparing these spectrograms, we observe that over-fitting phenomenon reduces the frequency of the formant and the fundamental frequency, which affects speech quality and speaker similarity. These figures suggest that the details of the frequency and the fluctuation of the formant become more abundant when the proposed method is used. This experiment shows that our method is effective in overcoming over-fitting.

\textbf{Prosody correlation}. To further verify the statistical significance of the expressiveness of each system, we extracted features related to the prosody: frame-level energy and lf0. We use 36 utterances to calculate the Pearson correlation coefficients between source audio and converted audio. The higher the Pearson correlation coefficient of the model, the higher the accuracy of the predicted prosodic attributes. As shown in Table~\ref{tab:pearson}, P3 gets the highest scores from the perspective of energy and lf0. This indicates that the prosody module can improve the performance of style transfer. The conclusion from objective measurement is line with the that from the subjective listening.

% \vspace{-15pt}
% \begin{table}[h]
% \centering
% \caption{Pearson correlation in energy and lf0.}
% \vspace{5pt}
% \label{tab:pearson}
% \begin{tabular}{{\centering}|m{1cm}<{\centering}m{1cm}<{\centering}m{1cm}<{\centering}m{1cm}<{\centering}m{1cm}}
% \toprule
%             & BL    & P1    & P2    & P3             \\ \hline
% Energy      & 0.765 & 0.727 & 0.729 & \textbf{0.755} \\ \hline
% Lf0         & 0.544 & 0.518 & 0.531 & \textbf{0.733} \\ \hline
% \end{tabular}
% \vspace{-15pt}
% \end{table}

% \vspace{-10pt}
\begin{table}[h] 
\centering
\caption{Pearson correlation in energy and lf0.}
\vspace{5pt}
\begin{tabular}{m{1cm}<{\centering}|m{1cm}<{\centering}m{1cm}<{\centering}m{1cm}<{\centering}m{1cm}}
\toprule
\label{tab:pearson}
\textbf{} & BL    & P1    & P2    & P3             \\ \midrule
Energy    & 0.765 & 0.727 & 0.729 & \textbf{0.755} \\ \hline
Lf0       & 0.544 & 0.518 & 0.531 & \textbf{0.733} \\ \bottomrule
\end{tabular}
\end{table}
\vspace{-15pt}

\section{CONCLUSION}

In this study, we propose a novel one-shot voice conversion framework for style transfer. This task is challenging as training on one utterance is easy to over-fit, which results in serious degradation of speaker similarity and style. To mitigate this challenge, we build on the recognition-synthesis framework and introduce a disentangled structure to explicitly model content, speaker identity and prosody which are originally entangled in speech. Specifically, we first adopt speaker normalization in content module to normalize speaker-related information. Furthermore, we add weight regularization during adaptation process to prevent over-fitting. Finally, to improve the expressiveness of converted speech, prosody module is added to one-shot voice conversion framework, which can extract rich prosody representation from source audio. Experimental results show that our proposed system outperforms several SOTA one-shot systems in terms of speaker similarity and style.

\ninept
\bibliographystyle{IEEE}
\bibliography{refs}

\end{document}